\documentclass[prd,aps,showpacs,,nofootinbib,superscriptaddress,preprintnumbers,epsf,psf]{revtex4}
\usepackage[english]{babel}
\usepackage{pstricks}
\usepackage[dvips]{graphicx}
\usepackage{epsf}

\begin{document}

\preprint{\hbox{RUB-TPII-08/09}}

\title{PHOTON-PHOTON COLLISION: AMBIGUITY AND DUALITY IN QCD FACTORIZATION THEOREM}

\author{{\slshape I.~V.~Anikin$^{1}$, I.~O.~Cherednikov$^{1,3,4}$,
N.~G.~Stefanis$^{2}$, O.~V.~Teryaev$^{1}$}\\[1ex]
$^1$Bogoliubov Laboratory of Theoretical Physics, JINR,
141980 Dubna, Russia\\
$^2$ Institut f\"{u}r Theoretische Physik II,
Ruhr-Universit\"{a}t Bochum, D-44780 Bochum, Germany\\
$^3$ INFN Cosenza, Universit$\grave{a}$
della Calabria, I-87036 Arcavacata di Rende, Italy\\
$^4$ ITPM, Moscow State University, RU-119899, Moscow, Russia}


\begin{abstract}
We discuss duality in ``two-photon''-like processes in the scalar
$\varphi^3_E$ model and also in the process
$\gamma^*\gamma\to\pi\pi$ in QCD.
Duality implies the equivalence between two distinct nonperturbative
mechanisms.
These two mechanisms, one involving a twist-$3$ Generalized
Distribution Amplitude, the other employing a leading-twist
Transition Distribution Amplitude, are associated with different
regimes of factorization.
In the kinematical region, where the two mechanisms overlap, duality
is observed for the scalar $\varphi^3_E$ model, while in the QCD case
the appearance of duality turns out to be sensitive to the particular
nonperturbative model applied and can, therefore, be used
as a tool for selecting the most appropriate one.
\end{abstract}
\maketitle

\section{Introduction}
\label{sec:intro}

The only known method today to apply QCD in a rigorous way is
based on the factorization of the dynamics and the isolation of a
short-distance part that becomes this way accessible to perturbative
techniques of quantum field theory
(see, \cite{Efremov-Radyushkin,Bro-Lep,Col-Sop-Ste89} and for
a review, for instance, \cite{Ste99} and references cited therein).
Then, the conventional systematic way of dealing with the long-distance
part is to parameterize it in terms of matrix elements of quark and
gluon operators between hadronic states (or the vacuum).
These matrix elements stem from nonperturbative phenomena and have to
be either extracted from experiment or be determined on the lattice.
In many phenomenological applications they are usually modeled
in terms of various nonperturbative methods or models.

Generically, the application of QCD to hadronic processes involves
the consideration of hard parton subprocesses and (unknown)
nonperturbative functions to describe binding effects.
Prominent examples are hard exclusive hadronic processes which
involve hadron distribution amplitudes (DAs), generalized
distribution amplitudes (GDAs), and generalized parton
distributions (GPDs) \cite{Diehl:2003,Bel-Rad,NonforRad,GPV}.
Applying such a framework, collisions of a real and a highly-virtual
photon provide a useful tool for studying a variety of fundamental
aspects of QCD.

\begin{figure}[h]
 \centerline{\includegraphics[width=0.4\textwidth,angle=0]{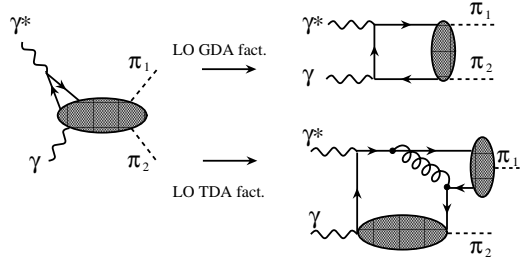}}
\vspace{0.0cm}
  \caption{Two ways of factorization: via the GDA mechanism
  and via the TDA mechanism.}
  \label{GDAvsTDA}
\end{figure}

Recently, nonperturbative quantities of a new kind were
introduced---transition distribution amplitudes
(TDAs) \cite{Frank-Pol,Pire-Szym,LPS06}---which are closely related to
the GPDs.
In contrast to the GDAs, the TDAs appear in the factorization procedure
when the Mandelstam variable $s$ is of the same order of magnitude
as the large photon virtuality $Q^2$, while $t$ is rather small.
Remarkably, there exists a reaction where both amplitude types,
GDAs and TDAs, can overlap.
This can happen in the fusion of a real and transversely polarized
photon with a highly-virtual longitudinally polarized photon, giving
rise to a final state which comprises a pair of pions.
The key feature of this reaction is that it can potentially follow
either path: proceed via twist-$3$ GDAs, or go through the
leading-twist TDAs, as illustrated in Fig.\ \ref{GDAvsTDA}.
Such an antagonism of alternative factorization mechanisms in this
reaction seems extremely interesting both theoretically and
phenomenologically and deserves to be studied in detail.

The intimate relation between these two mechanisms in the production
of a vector-meson pair was analyzed in \cite{PSSW} and it was found
that these mechanisms can be selected by means of the different
polarizations of the initial-state photon.
In contrast, for (pseudo)scalar particles, such as the pions, this
effect is absent enabling us to access the overlap region of both
mechanisms and their duality as opposed to their additivity.

In this talk, we will report on the possibility for duality
between these antagonistic mechanisms of factorization, associated
either with GDAs or with TDAs, in the regime where \textit{both}
Mandelstam variables $s$ and $t$ are rather small compared to the
large photon virtuality $Q^2$.

\section{Regimes of Factorization within the $\varphi^3_E$-model}
\label{sec:fact1}

Consider first the factorization of the scalar $\varphi^3_E$ model in
Euclidean space.
To study the four-particle amplitude in detail, it is particularly
useful to employ the $\alpha$-representation---see \cite{NonforRad}.
Then, the contribution of the leading ``box'' diagram can be written as
(while details can be found in \cite{An-Dual})
\begin{eqnarray}
\label{Amp1}
    {\cal A}(s,t,m^2)
    =-\frac{g^4}{16\pi^2}
    \int\limits_{0}^{\infty} \frac{\prod\limits_{i=1}^4 d\alpha_i}{D^2}
    \exp \biggl[ - \frac{1}{D} \left( Q^2 {\alpha_1\alpha_2}
    + s \alpha_2\alpha_4 +
    t {\alpha_1\alpha_3} + m^2 D^2 \right)\biggr],
\end{eqnarray}
where $m^2$ serves as a infrared (IR) regulator, $s>0$, $t>0$ are the
Mandelstam variables in the Euclidean region, and
$D=\sum\limits_{i=1}^4 \alpha_i$.
Assuming that $q^2=Q^2$ is large compared to the mass scale $m^2$
(which simulates here the typical scale of soft interactions), the
amplitude (\ref{Amp1}) can indeed be factorized.
As regards the other two kinematic variables $s$ and $t$, one can
identify three distinct regimes of factorization:
(a) $s\ll Q^2$ while $t$ is of order $Q^2$;
(b) $t\ll Q^2$ while $s$ is of order $Q^2$;
(c) $s,~t\ll Q^2$.

\textbf{Regime (a)}:
The process is going through the s-channel.
In this regime, the main contribution in the integral in
Eq.\ (\ref{Amp1}) arises from the integration over $\alpha_1$
when $\alpha_1\sim 0$:
\begin{eqnarray}
\label{GDA-alpha}
   {\cal A}_{\rm GDA}^{\rm as}(s,t,m^2)
    =-\frac{g^4}{16\pi^2}
    \int\limits_{0}^{\infty}
    \frac{d\alpha_2 \,d\alpha_3 \,d\alpha_4}{D^2_0} \
    \exp \left( - s \frac{\alpha_2\alpha_4}{D_0}- m^2 D_0 \right)
    \left[Q^2 \frac{\alpha_2}{D_0} + t \frac{\alpha_3}{D_0} + m^2
    \right]^{-1}\, .
\end{eqnarray}
Schematically this means that the propagator, parameterized by
$\alpha_1$, can be associated with the partonic (hard) subprocesses,
while the remaining propagator constitutes the soft part of the
considered amplitude, i.e., the scalar version of the GDA.

\textbf{Regime (b)}:
Here we have to eliminate from the exponential in Eq.\ (\ref{Amp1})
the variables $Q^2$ and $s$, which are large.
This can be achieved by integrating over the region $\alpha_2\sim 0$.
Performing similar manipulations as in regime (a), we find that the
scalar TDA amplitude can be related to the scalar GDA via
$
 {\cal A}_{\rm TDA}^{\rm as}(s, t, m^2)
=
 {\cal A}_{\rm GDA}^{\rm as}(t, s, m^2)
$.

\textbf{Regime (c)}:
The relevant regime to investigate duality is when it happens that
both variables $s$ and $t$ are simultaneously small compared to $Q^2$,
i.e., when $s,\, t \ll Q^2$.
In this case, there are two possibilities to extract the leading
$Q^2$-asymptotics, notably, we can either integrate over the region
$\alpha_1 \sim 0$, or integrate instead over the region
$\alpha_2 \sim 0$.
Clearly, these two options can be associated with (i) the GDA mechanism
of factorization with the meson pair scattered at a small angle in its
center-of-mass system or, alternatively, (ii) with the TDA mechanism of
factorization.
We stress that we may face double counting when naively adding these
two contributions.
We interpret such a behavior as a signal of an ingrained tendency for
duality between the GDA(s-channel) and the TDA (t-channel)
factorization mechanisms.

In order to verify the appearance of duality we carry out a numerical
investigation of the exact and the asymptotic amplitudes.
In doing so, we introduce the following ratios
$R_1={\cal A}_{\rm TDA}^{\rm as}/{\cal A}$ and
$R_2={\cal A}_{\rm GDA}^{\rm as}/{\cal A}$.
\begin{figure}[t]
 \centerline{\includegraphics[width=0.5\textwidth,angle=0]{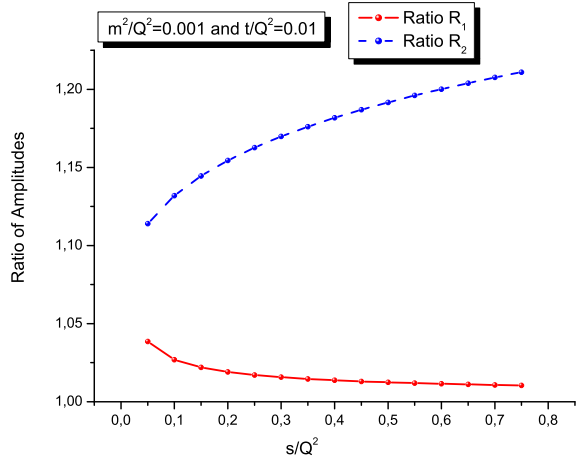}}
  \caption{The ratios $R_1$ and $R_2$ as functions of $s/Q^2$.}
  \label{fig-rat-1}
\end{figure}
Appealing to the symmetry of these ratios under the exchange of the
variables $s\leftrightarrow t$, we take $t/Q^2$ to be $0.01$ and
look for the variation of the ratios with $s/Q^2$.
This variation is illustrated in Fig.\ \ref{fig-rat-1} from which
one sees that in the region where $s/Q^2$ is rather small, i.e., in
the range $(0.01, \, 0.05 )$, both asymptotic formulae are describing
the exact amplitude with an accuracy of more than $90 \%$.
This behavior supports the conclusion that, when both Mandelstam
variables $s/Q^2$ and $t/Q^2$ assume values in the wide interval
$(0.001, \, 0.7)$, duality between the TDA and the GDA factorization
mechanisms emerges.

\section{TDA- and GDA-Factorizations for
         $\gamma\gamma^*\to\pi\pi$}
\label{sec:fact2}

Having discussed the appearance of duality between the GDA and the
TDA factorization schemes within a toy model, we now turn attention
to real QCD.
To analyze duality, we consider the exclusive $\pi^+\pi^-$ production
in a $\gamma_{\rm T}\gamma^{*}_{\rm L}$ collision, where the virtual
photon with a large virtuality $Q^2$ is longitudinally polarized,
whereas the other one is quasi real and transversely polarized.
Notice that the GDA and the TDA regimes correspond to the \textit{same}
helicity amplitudes.
Given that the considered process involves a longitudinally and a
transversally polarized photon, we are actually dealing with twist-3
GDAs \cite{AT-WW}.
On the other hand, for the twist-2 contribution, related to the meson
DA, we use the standard parametrization of the $\pi^+$-to-vacuum matrix
element which involves a bilocal axial-vector quark operator
\cite{Efremov-Radyushkin}.
Finally, the $\gamma\to\pi^-$ axial-vector matrix elements can be
parameterized in the form, cf.\ \cite{Pire-Szym},
\begin{eqnarray}
\langle \pi^-(p_2)| \bar \psi(-z/2)\gamma_\alpha \gamma_5
   [-z/2;z/2]\psi(z/2)
   |\gamma(q^\prime, \varepsilon^\prime)
   \rangle \stackrel{{\cal F}}{=}
   \frac{e}{f_\pi}\varepsilon^\prime_T
   \cdot\Delta_T P_\alpha A_1(x,\xi,t)\, ,
\label{eq:gpimeA}
\end{eqnarray}
where
$P=(p_2+q^\prime)/2$, and $\Delta=p_2-q^\prime$,
and noticing that the symbol $\stackrel{{\cal F}}{=}$ means Fourier
transformation and that the vector matrix element does not contribute
here.
To normalize the axial-vector TDA, $A_1$, we express it in terms of
the axial-vector form factor measured in the weak decay
$\pi\to l\nu_l \gamma$ \cite{PDG06,Byc08,An-Dual}.
The helicity amplitude associated with the TDA mechanism reads
\begin{eqnarray}
 {\cal A}^{\rm TDA}_{(0,j)}
=
 {\cal F}^{\rm TDA}
 \frac{\varepsilon^{\prime\,(j)}\cdot\Delta^T}{Q}
\end{eqnarray}
with
\begin{eqnarray}
\label{TDAhelam}
   {\cal F}^{\text{TDA}}=
   [4\,\pi\,\alpha_s(Q^2)]\frac{C_F}{2\,N_c}
   \biggl( {\rm tw-}2 \,\,\, {\rm DA}\biggr)
   \biggl( {\rm tw-}2 \,\,\, {\rm TDA}\biggr),
\end{eqnarray}
where
\begin{eqnarray}
   &&\biggl( {\rm tw-}2 \,\,\, {\rm DA}\biggr) =
   \int\limits_{0}^{1} dy \, \phi_\pi(y)
   \biggl( \frac{1}{y} + \frac{1}{\bar y} \biggr)\, ,
   \nonumber\\
   &&\biggl( {\rm tw-}2 \,\,\, {\rm TDA}\biggr)=
   \int\limits_{-1}^{1} dx \, A_1(x,\xi,t)\,
   \biggl( \frac{e_u}{\xi-x}-\frac{e_d}{\xi+x} \biggr),
   \label{eq:haTDA}
\end{eqnarray}
employing the 1-loop $\alpha_s(Q^2)$
in the $\overline{\rm MS}$-scheme with $\Lambda_{\rm QCD}=0.312$~GeV
for $N_f=3$ \cite{Kataev:2001kk}.
[Note that there is only a mild dependence on $\Lambda_{\rm QCD}$.]

Turning now to the helicity amplitude, which includes the
twist-$3$ GDA, we anticipate that it can be written as
(see, for example, \cite{AT-WW})
\begin{eqnarray}
{\cal A}_{(0,j)}^{\rm GDA}={\cal F}^{\rm GDA}
\frac{\varepsilon^{\,\prime\,(j)}\cdot\Delta^T}{Q}
\end{eqnarray}
with
\begin{eqnarray}
\label{GDAhelam}
{\cal F}^{\text{GDA}}=2 \frac{W^2+Q^2}{Q^2}
(e^2_u+e^2_d) \biggl( {\rm tw-3} \,\,{\rm GDA} \, {\rm WW}\biggr),
\end{eqnarray}
where
\begin{eqnarray}
\biggl( {\rm tw-3} \,\,{\rm GDA} \, {\rm WW}\biggr)=
\int\limits_{0}^{1} dy \, \partial_{\zeta} \Phi_1(y,\zeta,W^2)
\biggl( \frac{\ln{\bar y}}{y} - \frac{\ln{y}}{\bar y}\biggr)\, ,
\end{eqnarray}
with the partial derivative being defined by
$\partial_\zeta = \partial/\partial(2\zeta-1)$.
In deriving (\ref{GDAhelam}), we have used for the twist-$3$
contribution the Wandzura-Wilczek approximation.
Duality between expressions (\ref{TDAhelam}) and
(\ref{GDAhelam}) may occur in that regime, where both variables
$s$ and $t$ are simultaneously much smaller in comparison to the
large photon virtuality $Q^2$.
More insight into the relative weight of the amplitudes with TDA
or GDA contributions can be gained once we have modeled these
non-perturbative quantities.
We commence our analysis with the TDAs and, assuming a
factorizing ansatz for the $t$-dependence of the TDAs, we write
$A_1(x,\xi,t)=  2\frac{f_\pi}{m_\pi} \, F_{A}(t) A_1(x,\xi)$,
where the $t$-independent function $A_1(x,\xi)$ is normalized to
unity.
To satisfy the unity-normalization condition, we introduce a TDA
defined by
\begin{eqnarray}
 A_1(x, 1)
=
 \frac{A_1^{\rm non-norm}(x,1)}{\int\limits_{-1}^1 dx A_1^{\rm non-norm}(x,1)}
\end{eqnarray}
and continue with the discussion of the $t$-independent TDAs.
Recalling that we are mainly interested in TDAs in the region $\xi=1$
\cite{Efremov-Radyushkin,Bro-Lep}, it is useful to adopt the following
parametrization
\begin{eqnarray}
\label{TDAansatz}
    A_1^{\rm non-norm}(x, 1)
    =
      (1-x^2)\biggl( 1+ a_1 C^{(3/2)}_1(x)
    + a_2 C^{(3/2)}_2(x)+ a_4 C^{(3/2)}_4(x)\biggr) ,
\end{eqnarray}
where $a_1, \,a_2, \, a_4$ are free adjustable parameters, encoding
nonperturbative input, and the standard notations for Gegenbauer
polynomials are used.
It is not difficult to show that the TDA expressed by
Eq.\ (\ref{TDAansatz}) results from summing a $D$-term, i.e., the term
with the coefficient $a_1$, and meson-DA-like contributions.
For our analysis, we suppose that $a_1\equiv d_0$ \cite{GPV}, which is
equal to $-0.5$ in lattice simulations.
With respect to the parameters $a_2$ and $a_4$, we allow them to vary
in quite broad intervals, notably,
$a_2\in [0.3, \, 0.6]$ and $a_4\in [0.4, \, 0.8]$,
that would cover vector-meson DAs with very different
profiles at a normalization scale $\mu^2\sim 1\, {GeV^2}$
(see, for example, \cite{BM-rhomes}).
\begin{figure}[bt]
 \centerline{\includegraphics[width=0.5\textwidth,angle=0]{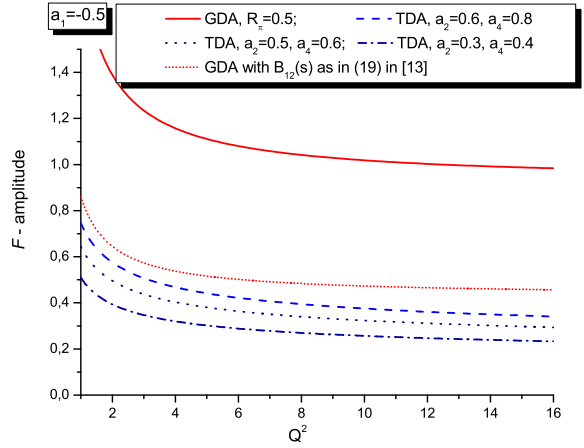}}
  \caption{Helicity amplitudes ${\cal F}^{\rm TDA}$ and
  ${\cal F}^{\rm GDA}$ as functions of $Q^2$, using $a_1=-0.5$
  found in lattice simulations.
  The value of $s/Q^2$ varies in the interval
  $[0.06,\, 0.3]$.
}
  \label{fig-TDA-GDA1}
\end{figure}
The function $\Phi_1(z,\zeta)$ is rather standard and well-known
(details in \cite{Diehl:2003, An-Dual}).

We close this section by summarizing our numerical analysis
presented in \cite{An-Dual}.
We calculated both functions
${\cal F}^{\rm TDA}$ and ${\cal F}^{\rm GDA}$,
and show the results in Fig.\ \ref{fig-TDA-GDA1}.
The dashed line corresponds to the function ${\cal F}^{\rm TDA}$,
where we have adjusted the free parameters to $a_2=0.6,\, a_4=0.8$.
The results, obtained for rather small values of these parameters, are
displayed by the broken lines in the same figure.
The dotted line denotes the function ${\cal F}^{\rm TDA}$ with
$a_2=0.5$ and $a_4=0.6$, whereas the dashed-dotted line employs
$a_2=0.3$ and $a_4=0.4$.
For comparison, we also include the results for ${\cal F}^{\rm GDA}$.
In that latter case, the dense-dotted line corresponds to the GDA
amplitude, where the expression for $\tilde B_{12}$ has been estimated
via Eq.\ (20) of \cite{An-Dual}, while the solid line represents the
simplest ansatz for $\tilde B_{12}$ with $R_\pi=0.5$.
From this figure one may infer that when the parameter $\tilde B_{12}$,
which parameterizes the GDA contribution, is estimated with the aid of
the Breit-Wigner formula (provided $s,\,t \ll Q^2$), there is duality
between the GDA and the TDA factorization mechanisms.
Hence, the model for $\Phi_1(z,\zeta)$, which takes into account the
corresponding resonances, can be selected by duality.

\section{Conclusions}
\label{sec:concl}

We have provided evidence that when both Mandelstam variables
$s$ and $t$ turn out to be much less than the large momentum scale $Q^2$,
with the variables $s/Q^2$ and $t/Q^2$ varying in the interval
$[0.001, \, 0.7]$, the TDA and the GDA factorization
mechanisms are equivalent to each other and operate in parallel.
We have also demonstrated that duality may serve as a tool for
selecting suitable models for the nonperturbative ingredients
of various exclusive amplitudes entering QCD factorization.
In this context, we observed that twist-3 GDAs appear to be dual
to the convolutions of leading-twist TDAs and DAs, multiplied by a QCD
effective coupling.

\section{Acknowledgments}

We would like to thank A.~P.~Bakulev, A.~V.~Efremov, N.~Kivel,
B.~Pire, M.~V.~Polyakov, M.~Prasza{\l}owicz, L.~ Szymanowski, and
S.~Wallon for useful discussions and remarks.
This investigation was partially supported by the Heisenberg-Landau
Programme (Grant 2008), the Alexander
von Humboldt Stiftung, the Deutsche Forschungsgemeinschaft under
contract 436RUS113/881/0, the EU-A7 Project \emph{Transversity},
the RFBR (Grants (grants 09-02-01149  and 07-02-91557),
the Russian Federation Ministry of Education and Science
(Grant MIREA 2.2.2.2.6546), the RF Scientific Schools grant 195.2008.9,
and INFN.



\end{document}